\documentclass[10pt,letterpaper]{article}
\usepackage{opex3}
\usepackage{wasysym}
\usepackage{SIunits}
\usepackage{cite}
\begin{document}

%%%%%%%%%%%%%%%%%% title page information %%%%%%%%%%%%%%%%%%
\title{Optical forces in nanowire pairs and metamaterials}

\author{Rongkuo Zhao,$^{1,2}$ Philippe Tassin,$^{1,3}$ Thomas Koschny,$^{1,4}$\\ and Costas M. Soukoulis$^{1,4}$}

\address{$^1$Ames Laboratory---U.S.~DOE,~and~Department~of~Physics~and~Astronomy,\\
             Iowa~State~University, Ames, Iowa~50011, USA}
\address{$^2$Applied Optics Beijing Area Major Laboratory, Department of Physics,\\
             Beijing Normal University, Beijing 100875, China}           
\address{$^3$Department~of~Applied~Physics~and~Photonics, Vrije~Universiteit~Brussel,\\
             Pleinlaan~2, B-1050~Brussel, Belgium}
\address{$^4$Institute~of~Electronic~Structure~and~Lasers~(IESL), FORTH, and~Department~of~Material Science~and~Technology,
             University~of~Crete, 71110~Heraklion, Crete, Greece}
\email{soukoulis@ameslab.gov}

\begin{abstract} 
We study the optical force arising when isolated gold nanowire pairs and metamaterials with a gold nanowire pair in the unit cell are illuminated with laser radiation. Firstly, we show that isolated nanowire pairs are subject to much stronger optical forces than nanospheres due to their stronger electric and magnetic dipole resonances. We also investigate the properties of the optical force as a function of the length of the nanowires and of the distance between the nanowires. Secondly, we study the optical force in a metamaterial that consists of a periodic array of nanowire pairs. We show that the ratio of the size of the unit cell to the length of the nanowires determines whether the electric dipole resonance leads to an attractive or a repulsive force, and we present the underlying physical mechanism for this effect.
\end{abstract}


%%%%%%%%%%%%%%%%%%%%%%% References %%%%%%%%%%%%%%%%%%%%%%%%%
%\bibliographystyle{osajnl}
%\bibliography{OpticalForce}

%%%%%%%%%%%%%%%%%%%%%%%%%%  body  %%%%%%%%%%%%%%%%%%%%%%%%%%
\section{Introduction}

The idea that electromagnetic radiation carries linear momentum goes back to James Clark Maxwell and his contemporaries~\cite{Maxwell-1873}; it is indeed the only way to reconcile Maxwell's equations with the principle of conservation of linear momentum~\cite{Jackson-1962}. Transfer of linear momentum from electromagnetic radiation to matter particles happens in the well-known phenomenon of radiation pressure, which is at least partially responsible for the tail of comets pointing away from the sun and is used in laser cooling to cool down atoms to temperatures close to the absolute zero~\cite{Chu-1998,Cohen-1998,Phillips-1998}. Recently, it has been proposed to harness optical forces, i.e., the forces that arise when linear momentum is transferred from photons to matter, in micro- and nanophotonic systems~\cite{Li-2008,VanThourhout-2010}.

Optical forces are commonly classified as either gradient or scattering forces. The gradient force is an optical force that is perpendicular to the propagation direction of the excitation electromagnetic field and is well known for its use in optical tweezers, where strong laser beams generate a piconewton force that is used for the manipulation of small dielectric particles, including DNA, enzymes, and biological entities such as cells and bacteria. The physics of optical tweezers can be understood by recognizing that a dielectric particle can lower its energy by moving towards a region with higher field intensity~\cite{Jackson-1962}. More recently, researchers have exploited the gradient optical force for all-optical actuation of nanomechanical systems. The typical setup consists of two suspended waveguides close to each other; each of the waveguides sits in the exponential tail of the other waveguide's mode and will therefore be subjected to an optical force of a few piconewtons per milliwatt of optical power~\cite{Povinelli-2005,Li-2008}. The force can be enhanced using various types of optical resonators, such as ring resonators, up to a few nanonewtons per milliwatt~\cite{Wiederhecker-2009}.

As opposed to the gradient force, the scattering optical force imparts momentum parallel to the propagation direction of the excitation field. It has been studied extensively in the field of cavity optomechanics. The typical setup in this case consist of an optical cavity of which one mirror is free to oscillate as a mechanical harmonic oscillator. It has been demonstrated theoretically and experimentally that the optical force on the mirror and the resulting coupling between the optical modes and the mechanical motion leads to peculiar effects such as amplification and cooling of the mechanical oscillator~\cite{Kippenberg-2008}. One of the goals of cavity optomechanics is to reduce the thermal occupancy of the mechanical oscillator such that the zero-point fluctuations of the---macroscopic---mechanical oscillator become observable. For our purposes, it is sufficient to note that the optical force can be strong enough to affect macroscopic masses.

All the systems discussed above need objects that are larger than the wavelength of light in order to generate a measurable optical force. In a study exploring the use of the optical force for the aggregation of metallic particles into periodic arrays, Hallock, Redmond, and Brus have shown that optical forces can also be obtained when subwavelength metallic spheres are illuminated by laser radiation~\cite{Hallock-2005}. When the laser frequency is chosen close to a collective plasmon resonance of the pair of metallic spheres, a much larger optical force can be excited due to plasmonic field enhancement~\cite{Chu-2007}. Early studies of optical forces on isolated nanowires include work by Rockstuhl and Herzig~\cite{Rockstuhl-2004} and by Halterman \emph{et al.}~\cite{Halterman-2005}.

\begin{figure}[bt!]
\begin{center}
\includegraphics{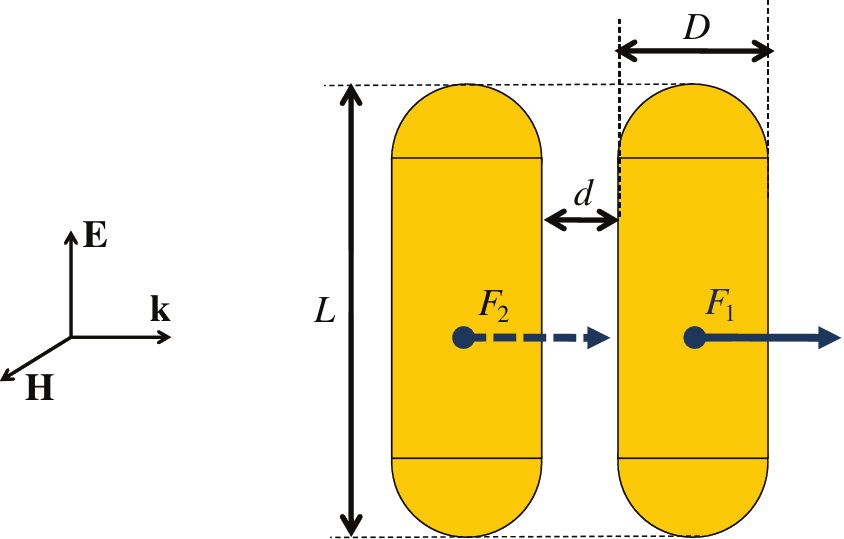}
\caption{The nanowire pair that is considered in this article for the purpose of studying the optical force in metamaterials. The propagation direction and polarization of the incident electromagnetic wave is indicated.}
\label{Fig:Nanowires}
\end{center}
\end{figure}

In this article, we want to explore optical forces in metamaterial elements. In recent years, researchers have found that strong currents can be excited in arrays of subwavelength electromagnetic resonators that are of quasistatic or plasmonic nature. Well-known examples of such metamaterial constituents are metal wires and split-ring resonators, i.e., metallic rings with a gap to provide sufficient capacitance. First fabricated in 2001~\cite{Shelby-2001}, metamaterials combining wires and split-ring resonators have been shown to produce media with negative permittivity and negative permittivity~\cite{Veselago-1968,Pendry-1998,Smith-2000}. Split-ring resonators and related structures are now well understood and many studies have been devoted to their optimization~\cite{Katsarakis-2004,Koschny-2004,Garcia-2005}. However, due to the saturation of the magnetic response above terahertz frequencies, they are not suitable for photonic systems. Here we will therefore use another prototype of a metamaterial element: metal nanowire pairs (see Fig.~\ref{Fig:Nanowires}). This structural element, which was independently proposed and fabricated by several groups~\cite{Zhang-2005,Dolling-2005,Shalaev-2005}, has been shown to exhibit a pronounced magnetic dipole resonance in the infrared and visible spectra.

We will continue this article with a description of the methods used to calculate the optical force in the considered structures: finite-element simulations and the Maxwell stress tensor formalism for the calculation of the optical force. Thereafter, in Sec.~\ref{Sec:Nanowire}, we will determine the optical force in an isolated nanowire pair illuminated by a laser beam. We investigate the strength of the force as a function of the geometric parameters of the nanowires and compare with the force between two spherical nanoparticles. Finally, in Sec.~\ref{Sec:Metamaterial}, we consider an array of nanowire pairs and we show that the interaction between neighbouring wire pairs can strongly modify the nature of the optical force, even if it influences the resonances of the individual wire pairs only slightly.

\section{Methods}\label{Sec:Methods}

We use COMSOL Multiphysics to solve Maxwell's equations for the scattering problem of an incident plane wave illuminating the nanowire pair or the array of nanowires. COMSOL Multiphysics uses the finite element method to discretize Maxwell's equations. For the isolated nanowire pairs, we embed the structure in a large spherical simulation domain that is terminated with a perfectly matched layer. For the nanowire pair arrays, we use a cuboid simulation domain with periodic boundary conditions on the faces parallel to the propagation direction and ports at the faces perpendicular to the propagation direction. Gold is modelled as a linear medium with complex, frequency-dependent permittivity obtained from experimental data by Johnson and Christy~\cite{Johnson-1972}.

After obtaining the full wave electromagnetic field solutions, we calculate the optical force on each of the nanowires using the Maxwell stress tensor formalism. The total force is the force on all charge carriers in the nanowire and is given by
\begin{equation}
F_i = \iiint_V{\left( \rho E_i + \epsilon_{ijk} J_j B_k \right) \mathrm{d}V},\label{Eq:Force}
\end{equation}
where $\epsilon_{ijk}$ is the Levi-Civita symbol and $V$ is a volume that fully encloses the nanowire. Since the metal is modelled by its complex permittivity, there are only bound charge carriers. Unfortunately, it is difficult (though not impossible) to implement Eq.~(\ref{Eq:Force}) in COMSOL Multiphysics, because of surface charge discontinuities at the edges of the metal wires. It is, nevertheless, straightforward to transform Eq.~(\ref{Eq:Force}) into~\cite{Jackson-1962}
\begin{equation}
F_i = -\epsilon_0\mu_0 \iiint_V{\frac{\partial S_i}{\partial t}\mathrm{d}V} + \iiint_V{\frac{\partial{T_{ij}}}{\partial{x_j}}\mathrm{d}V},\label{Eq:ForceMaxwellTimeDependent}
\end{equation}
where $S_i = \epsilon_0c^2\epsilon_{ijk} E_j B_k$ is Poynting's vector and $T_{ij}$ is the Maxwell stress tensor defined by
\begin{equation}
T_{ij} = \epsilon_0 \left( E_i E_j -\frac{1}{2}\delta_{ij}E^2 \right) + \frac{1}{\mu_0} \left( B_i B_j -\frac{1}{2}\delta_{ij}B^2 \right),
\end{equation}
with $\delta_{ij}$ the Kronecker delta. In the stationary harmonic regime employed in this article, the first term in the right-hand side of Eq.~(\ref{Eq:ForceMaxwellTimeDependent}) can be removed by averaging over an optical cycle; the second term can be transformed into a flux integral yielding
\begin{equation}
\left<F_i\right> = \oiint_S{\left<T_{ij}\right> n_j \mathrm{d}S}.\label{Eq:ForceMaxwellTimeAverage}
\end{equation}
Extending the integration surface to a box that is larger than the nanowire, all fields are well-defined and the integration can be straightforwardly carried out with COMSOL Multiphysics.

For the interpretation of the results, we will also make use of the electric and dipole moments of the current distributions in the nanowire pairs. We remind that all currents in the nanowires are modelled as polarization currents, so that both dipole moments can be obtained from the local electromagnetic fields available in COMSOL. The electric dipole moment is obtained from
\begin{equation}
p_\mathrm{z} = \iiint_V{P_\mathrm{z}\mathrm{d}V} = \iiint_V{\left( D_\mathrm{z} - \epsilon_0 E_\mathrm{z} \right)\mathrm{d}V}. \label{Eq:ElectricDipoleMoment}
\end{equation}
The magnetic moment is determined from $\mathbf{r}\times\mathbf{J}_\mathrm{P}$, yielding
\begin{equation}
m_\mathrm{y} = \iiint_V{\left( z \frac{\partial P_\mathrm{x}}{\partial t} - x \frac{\partial P_\mathrm{z}}{\partial t} \right)\mathrm{d}V}
             = -i\omega\iiint_V{\left[ z \left( D_\mathrm{x} - \epsilon_0 E_\mathrm{x} \right) 
                                     - x \left( D_\mathrm{z} - \epsilon_0 E_\mathrm{z} \right) \right]\mathrm{d}V}. \label{Eq:MagneticDipoleMoment}
\end{equation}

\section{Optical force in an isolated gold nanowire pair}\label{Sec:Nanowire}

\subsection{Physical nature of the optical force close to the resonances of a nanowire pair}

\begin{figure}[b!]
\begin{center}
\includegraphics{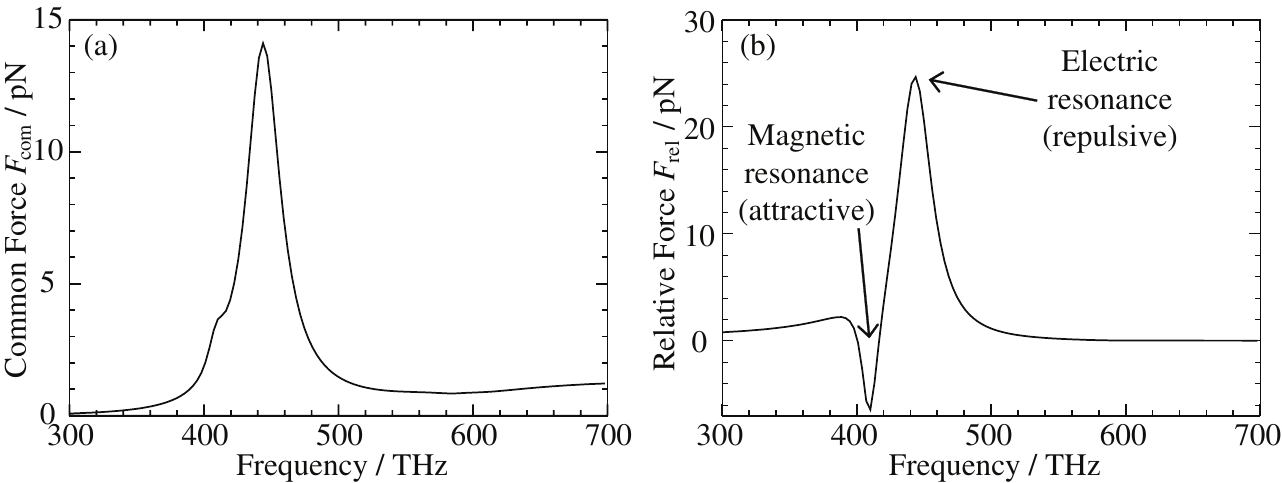}
\caption{(a) The common and (b) the relative optical force exerted on a pair of nanowires by an incident field of \unit{50}{\milli\watt\per\micro\meter\squared}. The length of the nanowires is $L = \unit{100}{\nano\meter}$, their diameter is $D = \unit{25}{\nano\meter}$, and the distance between the nanowires is $d = \unit{25}{\nano\meter}$. The relative force exhibits two resonances: the magnetic dipole resonance with an attractive force and the electric dipole resonance with a repulsive force.}
\label{Fig:IsolatedNW:Forces}
\end{center}
\end{figure}

\begin{figure}[t!]
\begin{center}
\includegraphics{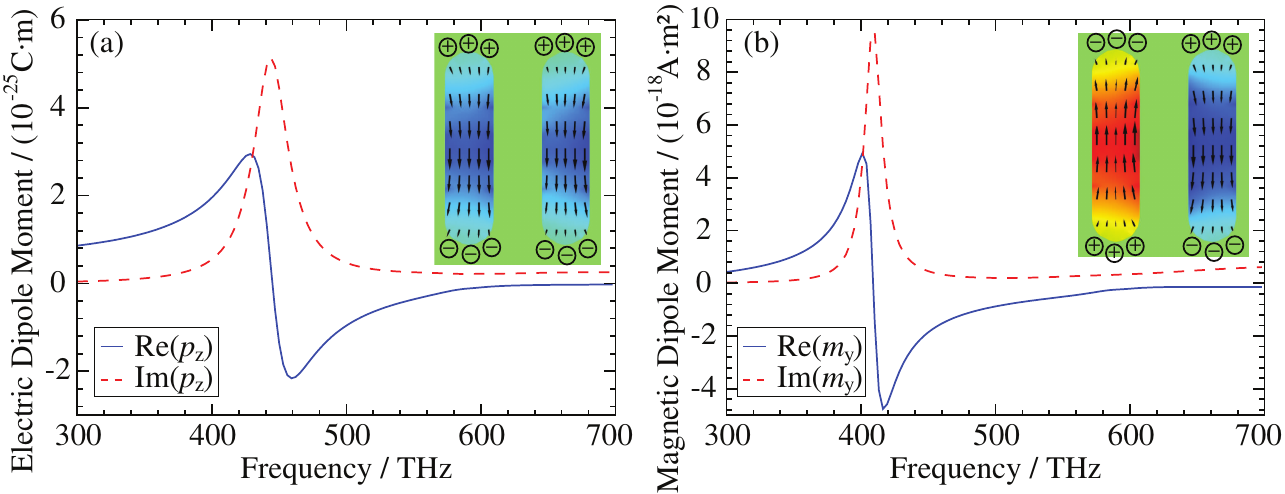}
\caption{Electric (a) and magnetic (b) dipole moment of the charge distribution in an isolated nanowire pair with length $L = \unit{100}{\nano\meter}$, diameter $D  = \unit{25}{\nano\meter}$, and interwire distance $D  = \unit{25}{\nano\meter}$. The insets show the charge and current distributions of the eigenmodes at \unit{445}{\tera\hertz} (electric dipole resonance) and at \unit{410}{\tera\hertz} (magnetic dipole resonance).}
\label{Fig:IsolatedNW:DipoleMoments}
\end{center}
\end{figure}

We start our study with an isolated gold nanowire pair with diameter of \unit{25}{\nano\meter}, length of \unit{100}{\nano\meter} and distance between the wires equal to \unit{25}{\nano\meter}; such nanowires can be fabricated with aspect ratios up to 7 by auto-catalytic growth from seeded metal salt solutions, identified by their particle plasmon resonances~\cite{Gom-2008}, and self-assembled side-by-side using molecular linkers~\cite{Park-2008}. The nanowire pair is illuminated with a plane wave with power density of \unit{50}{\milli\watt\per\micro\meter\squared} (see Fig.~\ref{Fig:Nanowires} for the direction of propagation and the polarization). Using the procedure described in Sec.~\ref{Sec:Methods}, we calculate the forces exerted on both nanowires: $F_1$ and $F_2$. For our purposes, it is more appropriate to consider the common force $F_\mathrm{com} = (F_1 + F_2)$ acting on the center of mass of the nanowire pair and the relative force $F_\mathrm{rel} = (F_1 - F_2)/2$ from which we can determine whether the optical forces will increase the separation between the nanowires (repulsive force) or decrease the separation (attractive force). In Fig.~\ref{Fig:IsolatedNW:Forces}(a)-(b), we plot both the common and relative forces for the \unit{100}{\nano\meter} long nanowire pair. We see that the relative force displays two pronounced resonances: an attractive force around \unit{410}{\tera\hertz} and a repulsive force at \unit{445}{\tera\hertz}. We can understand the nature of these forces from Fig.~\ref{Fig:IsolatedNW:DipoleMoments}(a)-(b), in which we plot the electric dipole moment and the magnetic dipole moment of the charge distribution in the nanowire pair. At the resonance close to \unit{410}{\tera\hertz}, the magnetic dipole moment is maximal and we can therefore identify this resonance as the antisymmetric eigenmode. The corresponding charge distribution, sketched in the inset of Fig.~\ref{Fig:IsolatedNW:DipoleMoments}(b), shows large opposite charges at the edges of the nanowires leading to an attractive Coulomb force. At the resonance close to \unit{445}{\tera\hertz}, the electric dipole moment exhibits a maximum, and this resonance therefore corresponds to the symmetric eigenmode with charges of equal sign at the edges of both nanowires [inset of Fig.~\ref{Fig:IsolatedNW:DipoleMoments}(a)], resulting in a repulsive Coulomb force. Note that the magnetic resonance is much narrower than the electric resonance. This is due to the weaker radiation damping of the magnetic dipole.

\subsection{Optical force as a function of the nanowires' length and the interwire distance}

\begin{figure}[t!]
\begin{center}
\includegraphics{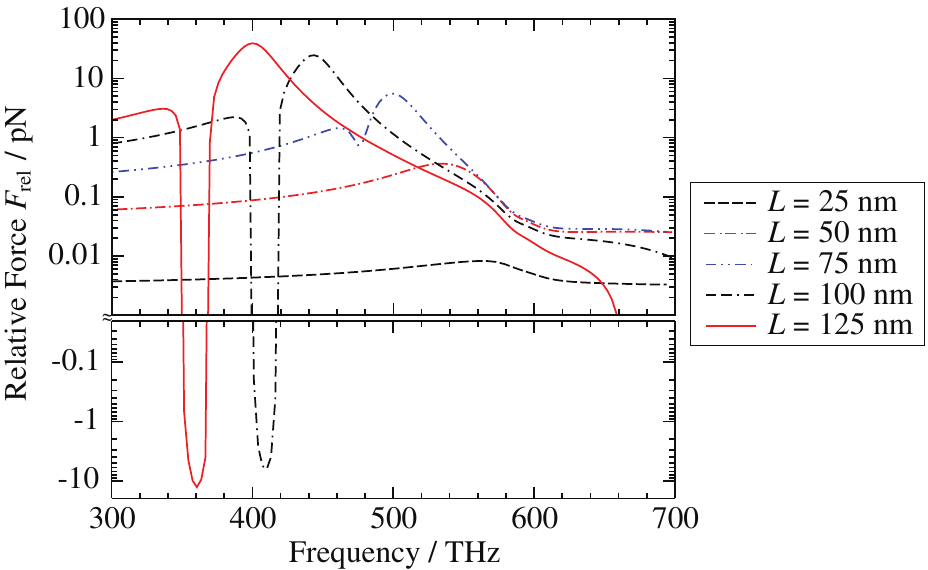}
\caption{The relative optical force in an isolated nanowire pair as a function of length. All nanowires have diameter $D = \unit{25}{\nano\meter}$ and interwire distance $d = \unit{25}{\nano\meter}$. Note that the force in the longer rods is exponentially larger than the force in the shortest sample (which has $L = D = \unit{25}{\nano\meter}$ and is therefore actually a pair of nanospheres).}
\label{Fig:IsolatedNW:Length}
\end{center}
\end{figure}

\begin{figure}[b!]
\begin{center}
\includegraphics{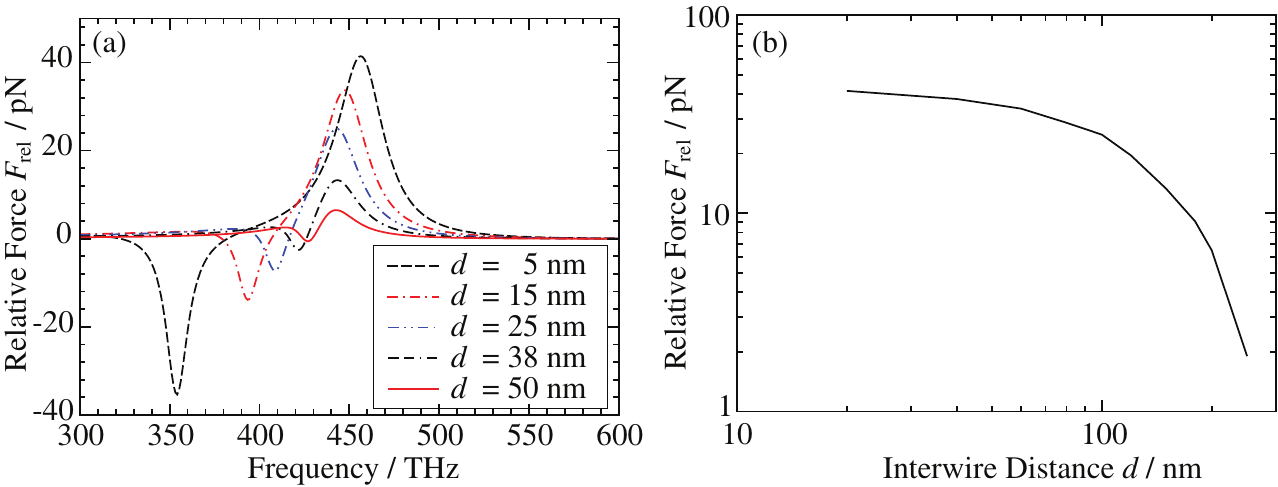}
\caption{(a)~The relative optical force in an isolated nanowire pair as a function of frequency for different interwire distances. All nanowires have diameter $D = \unit{25}{\nano\meter}$ and length $L = \unit{100}{\nano\meter}$. (b)~The relative optical force at the resonance frequency as a function of the interwire distance.}
\label{Fig:IsolatedNW:Distance}
\end{center}
\end{figure}

We continue our study with the question of how the strength of the optical force in nanowire pairs depends on the nanowires' length and the interwire distance. From the perspective of circuit theory, we can already anticipate that the force will become stronger with longer nanowires. Indeed, the larger inductance of a longer nanowire pair yields a higher quality factor of the circuit and, hence, larger charges at the ends of the nanowires. In addition, a longer nanowire pair will provide a stronger coupling to the incident field due to a stronger electromotive force. In order to test this hypothesis, we have plotted the average optical force exerted on nanowire pairs of different lengths in Fig.~\ref{Fig:IsolatedNW:Length}; we keep the nanowire diameter and the interwire distance fixed to \unit{25}{\nano\meter} and the power density of the incident wave is still \unit{50}{\milli\watt\per\micro\meter\squared}. We observe that the forces increase over several orders of magnitude when the length is increased from $L = \unit{25}{nm}$ to $L = \unit{125}{nm}$. Note that a nanowire with $L = \unit{25}{nm}$ is actually a sphere, which allows to compare the optical force between metallic spheres to the force between nanowires. We conclude that the optical force is significantly stronger in nanowire pairs. In future work, it will therefore be advantageous to make the nanowires as long as possible, of course within the limits of fabrication technology and the validity of the quasistatic theory employed in this article (i.e., the nanowires must remain smaller than the wavelength). Finally, we note that the attractive force disappears for nanorods shorter than $\unit{90}{\nano\meter}$; this is because the electric and magnetic resonances coincide for the chosen parameters and the electric dipole resonance dominates the charge distribution.

For completeness, we have also studied the optical force as a function of the interwire distance; the results are plotted in Figs.~\ref{Fig:IsolatedNW:Distance}(a)-(b). We observe that the overall peak value of the relative force increases with smaller interwire distance. This is in accordance with the explanation of the force in terms of the Coulomb force between the charges at the ends of the nanowires as explained above. The reader will note that the resonances frequencies shift, but this can again be readily explained with quasistatic circuit theory and is not the subject of this article.

\section{Optical force in a nanowire pair metamaterial}\label{Sec:Metamaterial}

\begin{figure}[b!]
\begin{center}
\includegraphics{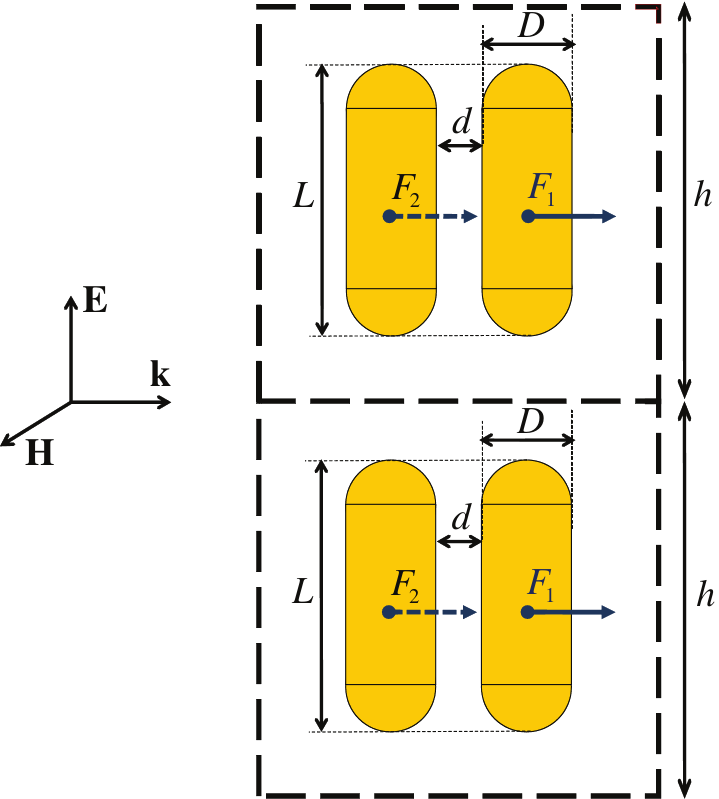}
\caption{Schematic of the array of nanowires studied in Sec.~\ref{Sec:Metamaterial}.}
\label{Fig:PeriodicNW:Array}
\end{center}
\end{figure}

We are now in good position to study a periodic array of nanowire pairs (see Fig.~\ref{Fig:PeriodicNW:Array} for an illustration). The effect of stacking metamaterials constituents in periodic arrays has been studied extensively in recent years. For most metamaterials, the coupling between neighbouring resonant element is merely renormalizing the circuit parameters (inductance, capacitance, and resistance) and although this coupling needs to be considered for the accurate design of metamaterials, the nature of the resonances remains unaltered~\cite{Zhou-2006}. Only when resonant elements are strongly coupled by placing them extremely close or under special conditions such as coupling to dark elements, hybridized resonances with novel behaviour can emerge, eventually leading to effects such as electromagnetically induced transparency~\cite{Gay-2002,Penciu-2008,EIT-2009,EIT-2009b,Zheludev-2008,Liu-2009}. Such extreme coupling is not considered here and one could therefore expect that the optical force is not significantly modified by the stacking of the nanowire pairs in a periodic array. Surprisingly, our simulations show that this is not true.

\begin{figure}[tb!]
\begin{center}
\includegraphics{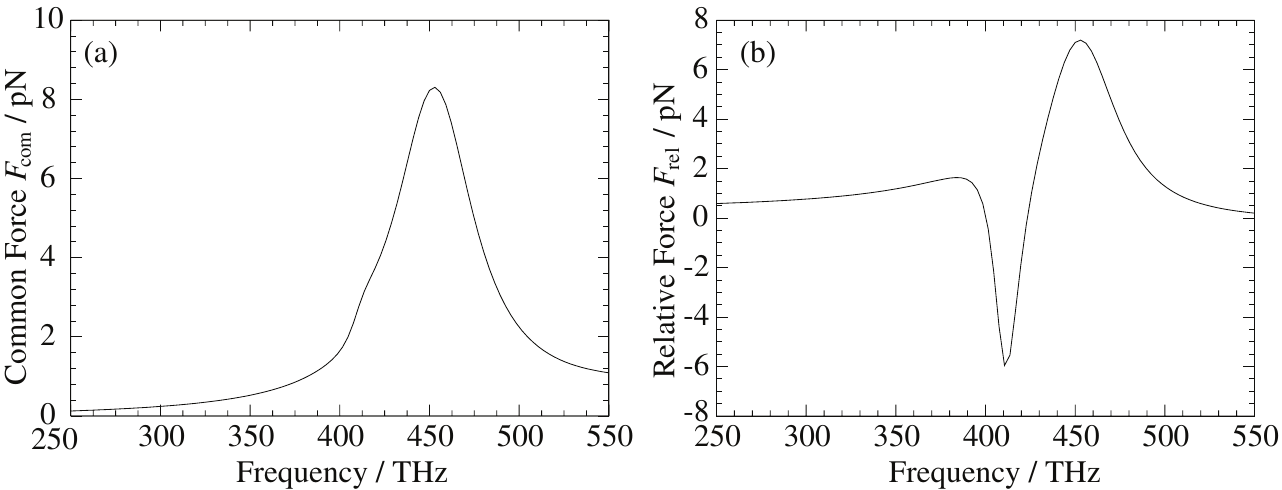}
\caption{(a) The common and (b) the relative optical force exerted on a pair of nanowires in a 2D periodic array with unit cell $\unit{100}{\nano\meter}\times\unit{200}{\nano\meter}$ when illuminated by a laser field of \unit{50}{\milli\watt\per\micro\meter\squared}. The length of the nanowires is $L = \unit{100}{\nano\meter}$, their diameter is $D = \unit{25}{\nano\meter}$, and the distance between the nanowires is $d = \unit{25}{\nano\meter}$. The relative force exhibits two resonances: the magnetic dipole resonance with an attractive force and the electric dipole resonance with a repulsive force.}
\label{Fig:PeriodicNW:Force}
\end{center}
\end{figure}

\begin{figure}[b!]
\begin{center}
\includegraphics{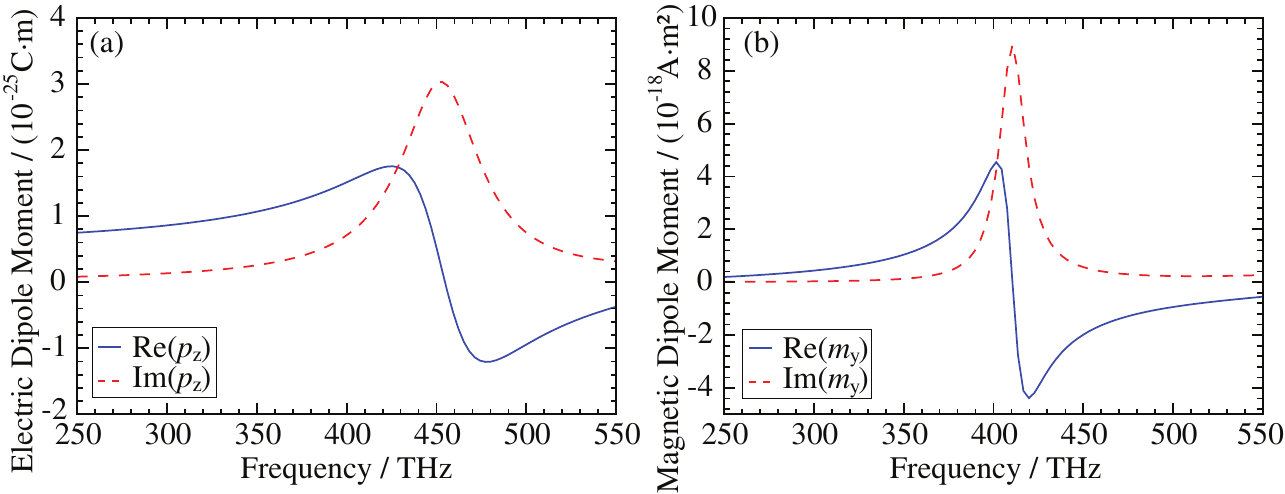}
\caption{Electric (a) and magnetic (b) dipole moment of the charge distribution in a 2D periodic nanowire pair metamaterial with unit cell $\unit{100}{\nano\meter}\times\unit{200}{\nano\meter}$. The length of the nanowires is $L = \unit{100}{\nano\meter}$, their diameter $D  = \unit{25}{\nano\meter}$, and the interwire distance equals $D  = \unit{25}{\nano\meter}$.}
\label{Fig:PeriodicNW:Moments}
\end{center}
\end{figure}

In Fig.~\ref{Fig:PeriodicNW:Force}(a)-(b), we plot the common and the relative optical forces on a \unit{100}{\nano\meter} long nanowire pair  with $D = \unit{25}{\nano\meter}$ and $d = \unit{25}{\nano\meter}$. The results look similar to the forces obtained for the isolated nanowire pair. First, we check that the nature of the resonances has remained; for the periodic array this can again be achieved by calculating the electric and magnetic dipole moments of the charge distribution in a single unit cell; the resulting dipole moments are shown in Fig.~\ref{Fig:PeriodicNW:Moments}(a)-(b). Although the stacking of the nanowire pairs in a periodic array does have some influence on its frequency response, the main features remain: we observe the magnetic resonance at \unit{410}{\tera\hertz}, so that the attractive force ($F_\mathrm{rel}<0$) at that frequency can still be attributed to the antisymmetric charge distribution associated with the magnetic dipole resonance as explained above for the isolated nanowire pair. Similarly, the electric dipole moment exhibits a resonance at \unit{450}{\tera\hertz}, so that the repulsive force around that frequency can be attributed to the symmetric charge distribution associated with electric dipole resonance.

\begin{figure}[tb!]
\begin{center}
\includegraphics{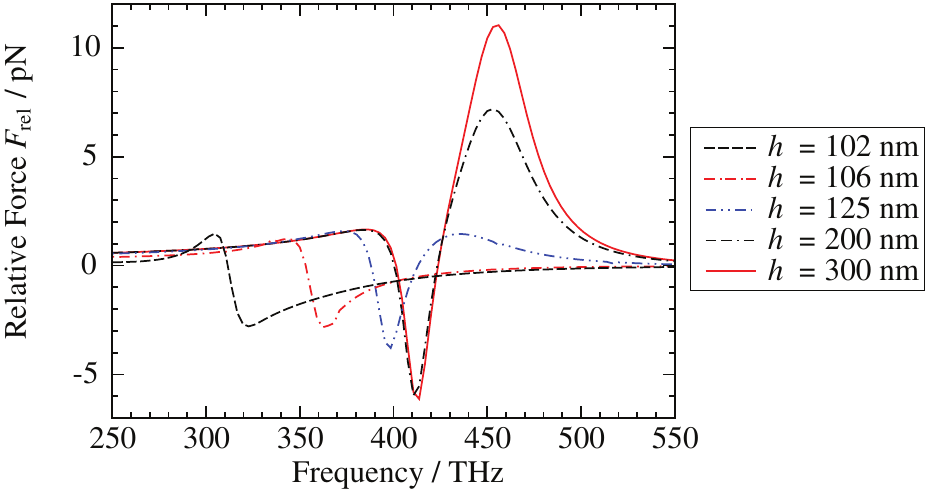}
\caption{The relative optical force in a periodic nanowire pair metamaterial as a function of the unit cell size height $h$ (for smaller unit cell height, the ends of the nanowires in neighbouring cells come closer). All nanowires have length $L = \unit{100}{\nano\meter}$, diameter $D = \unit{25}{\nano\meter}$, and interwire distance $d = \unit{25}{\nano\meter}$. The unit cell size is $\unit{100}{\nano\meter}\times h$. Note that the optical force decreases significantly for metamaterials with smaller unit cell height.}
\label{Fig:PeriodicNW:ForceVsUnitCell}
\end{center}
\end{figure}

We now determine the relative optical force on a nanowire pair in metamaterials with successively smaller unit cell height (the dimensions of the nanowire pairs and the width of the unit cell remain unaltered), so that the ends of the nanowires in neighbouring unit cells come closer; the results are displayed in Fig.~\ref{Fig:PeriodicNW:ForceVsUnitCell}. We start by noting that the positions of the resonances shift to lower frequencies. This behavior can be explained with quasistatic circuit theory as, for example, in Ref.~\cite{Zhou-2006}: the effective capacitance associated with both resonances is increased by the capacitance between two collinear nanowires in neighbouring unit cells, resulting in a lower resonance frequency.

More interestingly, however, is the strength of the optical force. We observe that the strength of the repulsive force is significantly diminished when the nanowire pairs are stacked in a periodic structure with smaller lattice constant. This cannot be explained based on our previous considerations, since circuit theory does not predict that the strength of the resonance would be changed so dramatically; this is confirmed by the electric and magnetic dipole moments (not shown here) that remain approximately constant at the resonance frequency. We can, however, understand what happens from the current distributions. If the distance between nanowires in adjacent unit cells becomes smaller than the distance between the nanowires of the same pair, the charges at the ends of two nanowires in adjacent unit cells start to cancel out and the charge distribution at the end of the nanowires will rather have the character of an electric dipole; the Coulomb force between the charges at nanowires' ends therefore start to satisfy an $1/r^3$ law typical for electric dipoles rather than the inverse square law found before and the resulting Coulomb force is significantly reduced.

\begin{figure}[tb!]
\begin{center}
\includegraphics{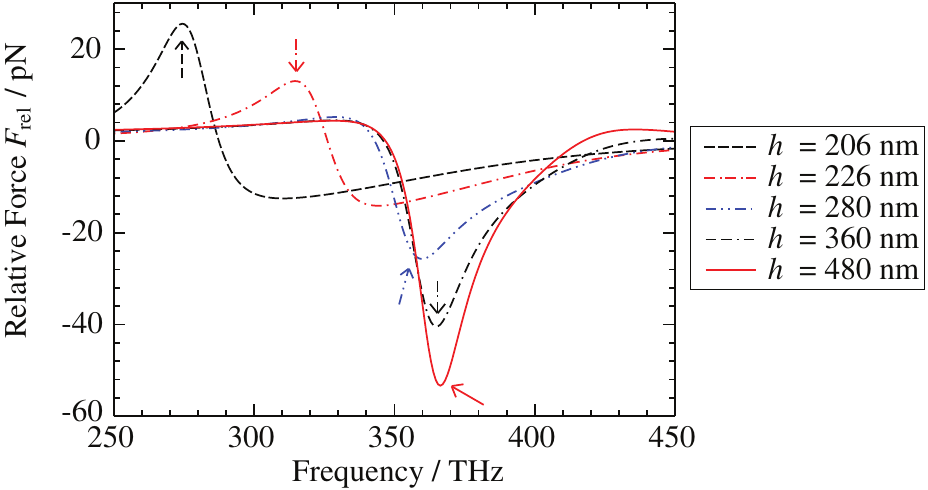}
\caption{The relative optical force in a periodic nanowire pair metamaterial as a function of the unit cell size height $h$ (for smaller unit cell height, the ends of the nanowires in neighbouring cells come closer). All nanowires have length $L = \unit{200}{\nano\meter}$, diameter $D = \unit{50}{\nano\meter}$, and interwire distance $d = \unit{50}{\nano\meter}$. The unit cell size is $\unit{200}{\nano\meter}\times h$. Note that the optical force reverses from attractive to repulsive for metamaterials with smaller unit cell height. The arrows mark the position of the magnetic dipole resonance.}
\label{Fig:PeriodicNW:ForceVsUnitCell2}
\end{center}
\end{figure}

\begin{figure}[tb!]
\begin{center}
\includegraphics{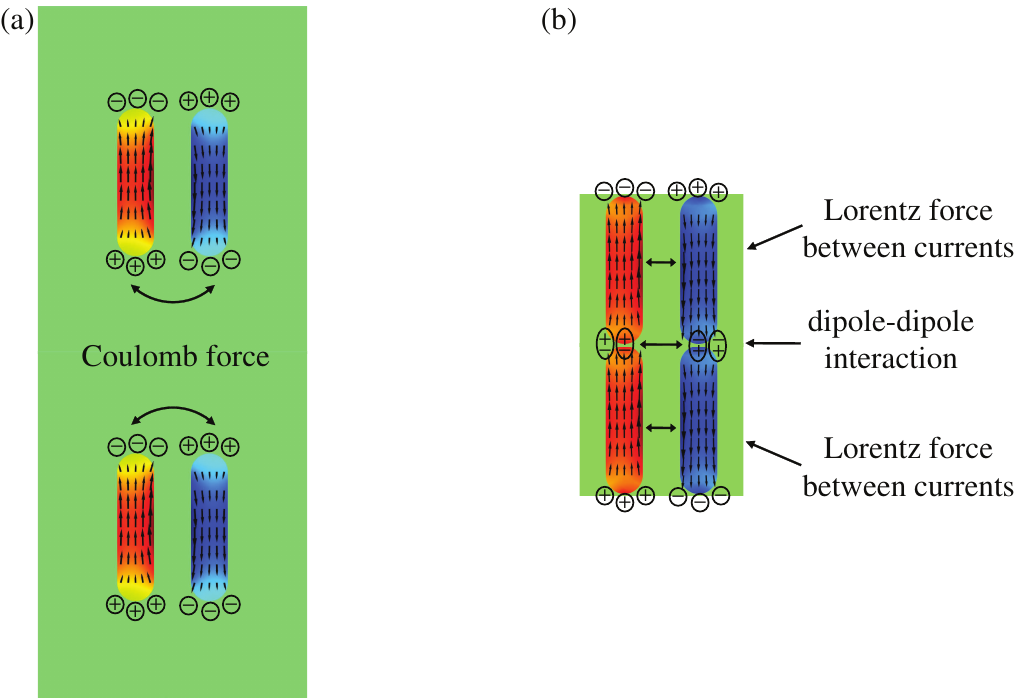}
\caption{Charge and current distribution in the nanowire pair metamaterial. (a)~If the height of the unit cell is substantially larger than the length of the nanowires, the force at the magnetic dipole resonance frequency has the nature of a Coulomb force between the charges at the ends of the wires and is attractive for the magnetic dipole resonance. (b)~If the height of the unit cell is only slightly larger than the length of the nanowires, the force at the magnetic dipole resonance frequency has the nature of an electric dipole-dipole interaction; the repulsive Lorentz force between the currents flowing in opposite directions can now be of comparable size or even dominate the optical force.}
\label{Fig:PeriodicNW:ChargeDistribution}
\end{center}
\end{figure}

For periodic arrays with larger nanowires, we find that the force associated with the magnetic dipole resonance can even be reversed from attractive to repulsive. In Fig.~\ref{Fig:PeriodicNW:ForceVsUnitCell2}, we display the relative force for a periodic array of nanowire pairs with length $L = \unit{200}{\nano\meter}$, diameter $D = \unit{50}{\nano\meter}$, and interwire distance $d = \unit{50}{\nano\meter}$. It can be shown from the dipole moments (indicated by arrows in Fig.~\ref{Fig:PeriodicNW:ForceVsUnitCell2}) that the large negative peak in the relative force for the samples with the largest unit cells ($h = \unit{480}{\nano\meter}$ and $h = \unit{360}{\nano\meter}$) corresponds to the magnetic dipole resonance. For the larger unit cells, the force is therefore attractive in agreement with the physical picture explained above. However, when we decrease the unit cell height, so that the ends of nanowires in neighbouring unit cells come closer, the force is first decreasing and finally reverses to become repulsive. It can again be shown from the dipole moments (indicated by arrows in Fig.~\ref{Fig:PeriodicNW:ForceVsUnitCell2}) that this positive peak is still associated with the magnetic dipole resonance, and not with the electric dipole resonance as one would expect based on the repulsive nature of the force. We repeat that the Coulomb force between the charges at the ends of the nanowires in one pair turns into a weaker dipole-dipole force when the unit cell height is decreased. In order to explain the reversion of the force from attractive to repulsive (see Fig.~\ref{Fig:PeriodicNW:ChargeDistribution} for an illustration)), we have to take into consideration the Lorentz force between the currents flowing in both nanowires. This force is repulsive, since the currents in both wires flow in opposite directions. The repulsive Lorentz force was negligible compared to the Coulomb force in all previously discussed samples, but now becomes first comparable and finally stronger than the attractive Coulomb force.

It is important that the reader keeps in mind that the strength of the underlying resonance remains approximately constant, i.e., that the magnetic (or electric) dipole moment depends only slightly on the unit cell size. The examples above therefore show that the mere action of stacking a metamaterial element in a periodic array can dramatically change the optical force, even though the underlying quasistatic resonances are not altered significantly.

\section{Conclusions}

We have presented a study of the optical force in metamaterials. In the first part of this article, we have shown that appreciable optical forces can be generated in gold nanowire pairs when illuminated with laser radiation. Specifically, we have shown that gold nanowire pairs allow for optical forces that are significantly larger than those observed in pairs of nanospheres~\cite{Hallock-2005,Chu-2007}. We have presented a complete physical picture explaining the forces as Coulomb forces between the charges accumulated at the ends of the nanowires. Repulsive forces originate from the electric dipole resonance of the nanowire pair and attractive forces originate from the magnetic dipole resonance.

In the second part of the article, we have considered a metamaterial consisting of a periodic array of nanowire pairs. As long as the unit cell size is large enough---and the nanowire pairs in neighbouring unit cells are well separated---we recover the same behavior as for the isolated nanowire pairs. However, with closer stacking of the nanowire pairs, we find that the nature of the optical force is altered dramatically, even though the underlying electric and magnetic dipole resonances are modified only slightly. In some nanowire pair geometries, it is even possible to turn the optical force associated with the magnetic dipole resonance from attractive to repulsive. Two effects contribute to such behavior: (i) As the distance between nanowires in neighbouring unit cells becomes smaller than the distance between nanowires in the same unit cell, the optical force changes from a Coulomb force between point charges into a much weaker dipole-dipole interaction. (ii) As the force between the charges at the ends of the nanowires becomes smaller, the Lorentz force between the currents in the nanowires will finally start to dominate, which explains the reversion from an attractive to a repulsive force.

\section*{Acknowledgments}

This work was supported by the AFOSR under MURI Grant No.\ FA9550-06-1-0337. Work at Ames Laboratory was partially supported by the Department of Energy (Basic Energy Sciences) under Contract No.\ DE-AC02-07CH11358. Work at VUB was supported by BelSPO (Grant No.\ IAP6/10). R.~Z.\ acknowledges the China Scholarship Council (CSC) for financial support. P.~T.~is supported by a Fellowship of the Belgian American Educational Foundation.

\end{document}